\documentclass[twocolumn]{aastex62}

\newcommand{\bd}{\begin{displaymath}}
\newcommand{\ed}{\end{displaymath}}
\newcommand{\be}{\begin{equation}}
\newcommand{\ee}{\end{equation}}
\newcommand{\beaa}{\begin{eqnarray*}}
\newcommand{\eeaa}{\end{eqnarray*}}
\newcommand{\bea}{\begin{eqnarray}}
\newcommand{\eea}{\end{eqnarray}}

\newcommand\HST{\textit{HST}}

\submitjournal{ApJ}

\shorttitle{On the accuracy of time-delay cosmography with supernova Refsdal}
\shortauthors{Grillo et al.} 

\begin{document}

\title{On the accuracy of time-delay cosmography in the Frontier Fields Cluster MACS J1149.5+2223 with supernova Refsdal}

\correspondingauthor{Claudio Grillo}
\email{claudio.grillo@unimi.it}

\author[0000-0002-5926-7143]{C.~Grillo}
\affiliation{Dipartimento di Fisica, Universit\`a  degli Studi di Milano, via Celoria 16, I-20133 Milano, Italy}
\affiliation{Dark Cosmology Centre, Niels Bohr Institute, University of Copenhagen, Lyngbyvej 2, DK-2100 Copenhagen, Denmark}
\affiliation{INAF - IASF Milano, via A. Corti 12, I-20133 Milano, Italy}

\author{P.~Rosati}
\affiliation{Dipartimento di Fisica e Scienze della Terra, Universit\`a degli Studi di Ferrara, Via Saragat 1, I-44122 Ferrara, Italy}
\affiliation{INAF - Osservatorio Astronomico di Bologna, via Gobetti 93/3, I-40129 Bologna, Italy}

\author{S.~H.~Suyu}
\affiliation{Max-Planck-Institut f{\"u}r Astrophysik, Karl-Schwarzschild-Str.~1, 85748 Garching, Germany}
\affiliation{Physik-Department, Technische Universit{\"a}t M{\"u}nchen, James-Franck-Straße 1, 85748 Garching, Germany}
\affiliation{Academia Sinica Institute of Astronomy and Astrophysics (ASIAA), 11F of ASMAB, No.1, Section 4, Roosevelt Road, Taipei 10617, Taiwan}

\author{G.~B.~Caminha}
\affiliation{Kapteyn Astronomical Institute, University of Groningen, Postbus 800, 9700 AV Groningen, the Netherlands}

\author{A.~Mercurio}
\affiliation{INAF - Osservatorio Astronomico di Capodimonte, Via Moiariello 16, I-80131 Napoli, Italy}

\author{A.~Halkola}



\begin{abstract}

  We study possible systematic effects on the values of the cosmological parameters defining the expansion rate and the geometry of the universe measured through strong lensing analyses of the Hubble Frontier Fields galaxy cluster MACS J1149.5+2223. We use the observed positions of a large set of spectroscopically selected multiple images, including those of supernova ``Refsdal'' with their estimated time delays. Starting from our reference model in a flat $\Lambda$CDM cosmology, published in \citet{gri18}, we confirm the relevance of the longest measurable time delay, between SX and S1, and an approximately linear relation between its value and that of $H_{0}$. We perform true blind tests by considering a range of time delays around its original estimate of $345 \pm 10$ days, as an accurate measurement of this time delay has not been published to date and it is not known at the time of writing. We investigate separately the impact of a constant sheet of mass at the cluster redshift, of a power-law profile for the mass density of the cluster main halo and of some scatter in the cluster member scaling relations. Remarkably, we find that these systematic effects do not introduce a significant bias on the inferred values of $H_{0}$ and $\Omega_{\rm m}$ and that the statistical uncertainties dominate the total error budget: a 3\% uncertainty on the time delay of image SX translates into approximately 6\% and 40\% (including both statistical and systematic $1\sigma$) uncertainties for $H_{0}$ and $\Omega_{\rm m}$, respectively. 
    Furthermore, our model accurately reproduces the extended surface brightness distribution of the supernova host, covering more than $3 \times 10^{4}$ \HST\ pixels.
  We also present the interesting possibility of measuring the value of the equation-of-state parameter $w$ of the dark energy density, currently with a 30\% uncertainty. We conclude that time-delay cluster lenses, when characterized with sufficient spectroscopic data, have the potential to become soon an alternative and competitive cosmological probe.

\end{abstract}



\section{Introduction}
\label{sec:intro} 
After nearly a century from its first estimates \citep{lemaitre27,hub29}, the exact value of the cosmic expansion rate, the Hubble constant $H_{0}$, is still hotly debated.
The most recent measurements of $H_{0}$, based on the  distance ladder from the SHOES program \citep{rie19} and from the \emph{Planck} satellite \citep{pla18}, cannot be reconciled. Accurate and precise estimates from independent methods and calibrations (e.g., \citealt{abb17}; \citealt{freedman19}; \citealt{reid13}; \citealt{won19}) are thus crucial to clarify whether an extention of the standard cosmological model is needed. Time delays between the multiple images of supernovae (SNe) and quasars (QSOs) strongly lensed by galaxy clusters can place interesting new constraints on the values of $H_{0}$ and of the parameters defining the global geometry of the Universe \citep[e.g.,][]{Refsdal64, Chen19, won19, Shajib19}.

In this paper, we use a full strong lensing model of the Hubble Frontier Fields (HFF) galaxy cluster MACS J1149.5$+$2223 (hereafter MACS 1149; $z = 0.542$), where the first multiply-imaged and spatially-resolved SN ``Refsdal'' ($z = 1.489$) was discovered \citep{kel15,kel16a}. We extend our previous analysis (\citealt{gri18}, hereafter G18) and show that a large set of spectroscopically confirmed multiple images from different sources (\citealt{tre15b}; \citealt{gri16}, hereafter G16) and the measured time delays between the multiple images of a variable source (\citealt{kel16}; \citealt{rod16}) in a lens galaxy cluster can provide accurate measurements of the values of the Hubble constant and of cosmological parameters, which are competitive with those from other cosmological probes. Alternative methods exploiting SN Refsdal for cosmography were presented in other studies (\citealt{veg18}; \citealt{wil19}).

\section{Methods}
\label{sec:met}

Our reference strong lensing model of MACS 1149 (labeled as r/REF in the following tables and figures) is the one detailed in G16 and G18. We refer the reader to those papers for a comprehensive decription. Here, we summarize briefly its main caracteristics. To reconstruct the cluster total mass distribution and infer the values of the considered cosmological parameters, we use the observed positions of 89 multiple images from 28 different point-like sources, with redshifts between 1.240 and 3.703, and the values of the time delays of the images S2, S3, and S4, relative to S1, of SN Refsdal, as measured by using polynomial fitting by \citet{rod16}. The value of the SX-S1 time delay (${\rm \Delta t_{\rm SX:S1}}$) is based on the first results by \citet{kel16} and is varied between 315 and 375 days, with a constant uncertainty of 10 days. The cluster total mass distribution is modeled with a combination of 3 cored elliptical pseudo-isothermal and 300 dual pseudo-isothermal mass density profiles for, respectively, the extended dark-matter halos and the galaxy members of the cluster. Flat ($\Omega_{\rm m}+\Omega_{\Lambda} = 1$) $\Lambda$CDM models with uniform priors on the values of $H_{0}$, between 20 and 120 km~s$^{-1}$ Mpc$^{-1}$, and $\Omega_{\rm m}$, between 0 and 1, are considered. The software {\sc Glee} \citep{SuyuHalkola10,SuyuEtal12} is used to conduct the entire strong-lensing study.

First, to analyze the dependence of the inferred value of $H_{0}$ on the observed value of the time delay between SX and S1, we perform full statistical analyses with the reference model and nine different values for this time delay (i.e., 315, 322, 330, 337, 345, 353, 360, 368 or 375 days), keeping its uncertainty fixed to 10 days. We also check the importance of the SX-S1 time delay in estimating the values of $H_{0}$ and $\Omega_{\rm m}$ by comparing the results of the reference model with and without this observable quantity.

Then, to test the effect of the most important sources of systematic uncertainties on the measurement of the values of the cosmological parameters, we consider additional models, where we (1) add to the reference model (with 315, 345 or 375 $\pm10$ days for the SX-S1 time delay) a constant sheet of mass at the cluster redshift with a uniform prior between $-$0.2 and 0.2 on the value of the convergence k$_{0}$ (models labeled as $\kappa$); (2) substitute (with 315, 345 or 375 $\pm10$ days for the SX-S1 time delay) the pseudo-isothermal profile (i.e., $\rho(r) \sim r^{-2}$, in the outer regions $r \gg r_{\rm c}$) of the cluster main halo with a more general power-law profile (i.e., $\rho(r) \sim r^{-\gamma}$, in the outer regions $r \gg r_{\rm c}$) with a uniform prior between 1.4 and 2.6 on the value of $\gamma$ (models labeled as $\gamma$); (3) use five different realizations (all with 345$\pm10$ days for the SX-S1 time delay) of the values of the Einstein angles ($\vartheta_{\mathrm{E}}$) and truncation radii ($r_{\mathrm{t}}$) of the cluster members (models labeled as s1-s5), randomly scattered with normal distributions centered on the values obtained with total mass-to-light ratios increasing according to the galaxy near-IR luminosities and with standard deviations equal to 10\% of their values  (see \citealt{ber19} for recent results on cluster member modeling with stellar kinematics information).

Next, to quantify the goodness of our best-fitting reference model (with a time delay between SX and S1 of 345 $\pm$ 10 days), we compare the observed and model-predicted surface brightness distribution of the multiple images of the SN Refsdal host. We note that in this model all the sources are approximated as point-like objects and the extended surface brightness information has not been exploited yet. 
 We then use the best-fitting reference model, based on the point images, to reconstruct the SN host galaxy on a grid of pixels on the source plane from the lensed and distorted multiple images of the SN host galaxy.
To suppress the contaminating contribution of the cluster members' 
light to the lensed SN host galaxy light,
we use a linear combination of the \HST\ F606W and F435W bands, with a pixel size of 0.06\arcsec, from the HFF project. Due to the very large number of pixels over which the multiple images of the SN Refsdal host galaxy extend, our computing memory sets a maximum size of $75 \times 75$ pixels for the reconstruction grid of the source surface brightness. From the reconstructed SN host galaxy surface brightness on the source plane, we then lens it with the mass model to the image plane to compare with the observed lensed images.

Finally, we consider models (with 315, 345 or 375 $\pm10$ days for the SX-S1 time delay) in a flat ($\Omega_{\rm m}+\Omega_{\Lambda} = 1$) $w$CDM cosmology, in which the dark energy density is time dependent, with an equation-of-state parameter $w$. Here, the value of $w$ is also free to vary between $-2$ and $0$, with a uniform prior.

\section{Results}
\label{sec:res}

We remark that all the results presented here come from a blind analysis, because the exact value of the time delay ${\rm \Delta t_{\rm SX:S1}}$ was not known at the time of analysis and writing.

\begin{figure}
\centering
\includegraphics[width=0.49\textwidth]{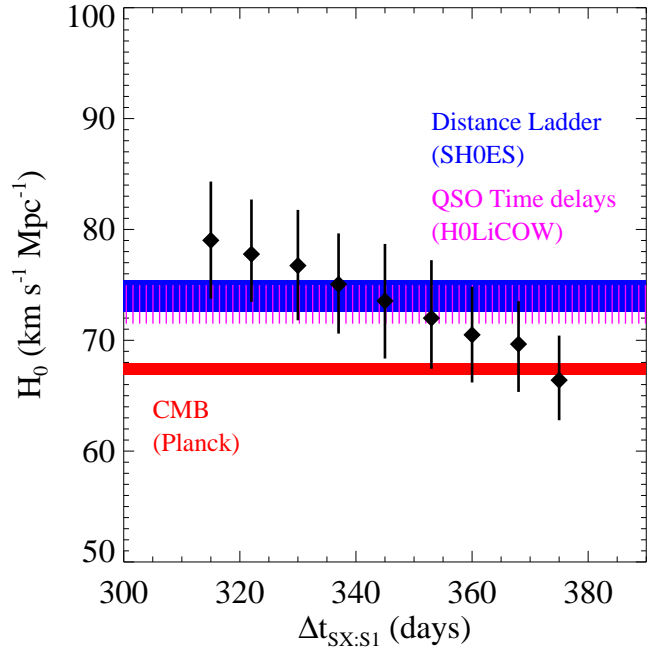}
\caption{Inferred median values (diamonds) and $68\%$ CL intervals (bars) of $H_{0}$ for reference (r) models with different adopted values for the time delay between SX and S1 (315, 322, 330, 337, 345, 353, 360, 368 or 375 days) and a constant uncertainty of 10 days. Flat $\Lambda$CDM models ($\Omega_{\rm m}+\Omega_{\Lambda} = 1$) with uniform priors on the values of the cosmological parameters ($H_{0} \in [20,120]$ km~s$^{-1}$~Mpc$^{-1}$ and $\Omega_{\rm m} \in [0,1]$) are considered. Final MCMC chains have $2 \times 10^{6}$ samples for each model. The blue, purple and red bands indicate, respectively, the credible intervals, at 1$\sigma$ CL, from SH0ES (\citealt{rie19}), H0LiCOW (\citealt{won19}) and Planck (\citealt{pla18}).}
\label{fig2}
\end{figure}

\begin{table}
\centering
\caption{Median values and 1$\sigma$ CL uncertainties of the Hubble constant $H_{0}$ (in km s$^{-1}$ Mpc$^{-1}$) and $\Omega_{\rm m}$ for the models shown in Figure \ref{fig1}.}
\begin{tabular}{cccc}
\hline\hline \noalign{\smallskip}
${\rm \Delta t_{\rm SX:S1}}$ & model & $H_{0}$ & $\Omega_{\rm m}$ \\
\noalign{\smallskip} \hline \noalign{\smallskip}
& r & $79.0 \pm 5.3$ & $0.32 \pm 0.14$ \\
315d & $\kappa$ & $79.7 \pm 5.2$ & $0.39 \pm 0.16$ \\
& $\gamma$ & $78.7 \pm 4.8$ & $0.33 \pm 0.12$ \\
\noalign{\smallskip} \hline
& r & $73.5 \pm 5.2$ & $0.37 \pm 0.16$ \\
345d & $\kappa$ & $73.1 \pm 4.3$ & $0.38 \pm 0.15$ \\
& $\gamma$ & $73.9 \pm 3.9$ & $0.39 \pm 0.13$ \\
\noalign{\smallskip} \hline
& r & $66.4 \pm 3.8$ & $0.48 \pm 0.18$ \\
375d & $\kappa$ & $66.4 \pm 4.0$ & $0.49 \pm 0.21$ \\
& $\gamma$ & $65.5 \pm 3.7$ & $0.47 \pm 0.10$ \\
\noalign{\smallskip} \hline
& s1 & $73.5 \pm 4.3 $ & $0.37 \pm 0.15$ \\
& s2 & $73.5 \pm 4.7$ & $0.40 \pm 0.15$ \\
345d & s3 & $72.3 \pm 4.3$ & $0.34 \pm 0.13$ \\
& s4 & $73.5 \pm 4.2$ & $0.36 \pm 0.15$ \\
& s5 & $72.9\pm 4.3$ & $0.38 \pm 0.17$ \\
\noalign{\smallskip} \hline
\end{tabular}
\label{stat}
\end{table}

We show in Figure \ref{fig2} the median values of $H_{0}$ and the $68\%$ confidence level (CL) intervals obtained by varying the value of the time delay between SX and S1 from 315 to 375 days, with linear steps of 7-8 days, as described in Sect. \ref{sec:met}. The inferred value of $H_{0}$ decreases from 79.0 to 66.4 km s$^{-1}$ Mpc$^{-1}$ as the time-delay value increases. The assumed constant uncertainty of 10 days on the SX-S1 time delay translates into (1$\sigma$) statistical errors for $H_{0}$ between 5.8 and 7.0\%, with a median value of 6.1\%, consistent with the results presented in G18. 

We have also checked that an error of 2\% on the value of ${\rm \Delta t_{\rm SX:S1}} = 345$ days (i.e., 7 days) results into an approximately 1\% smaller error on the value of $H_{0}$ (similarly to G18) and that a smaller uniform prior between 0.05 and 0.5 for the value of $\Omega_{\rm m}$ (as adopted, for instance, in \citealt{won19}) reduces by more than 10\% the error on its value and by less than 1\% the error on the value of $H_{0}$.

\begin{figure*}
\centering
\includegraphics[width=0.4\textwidth]{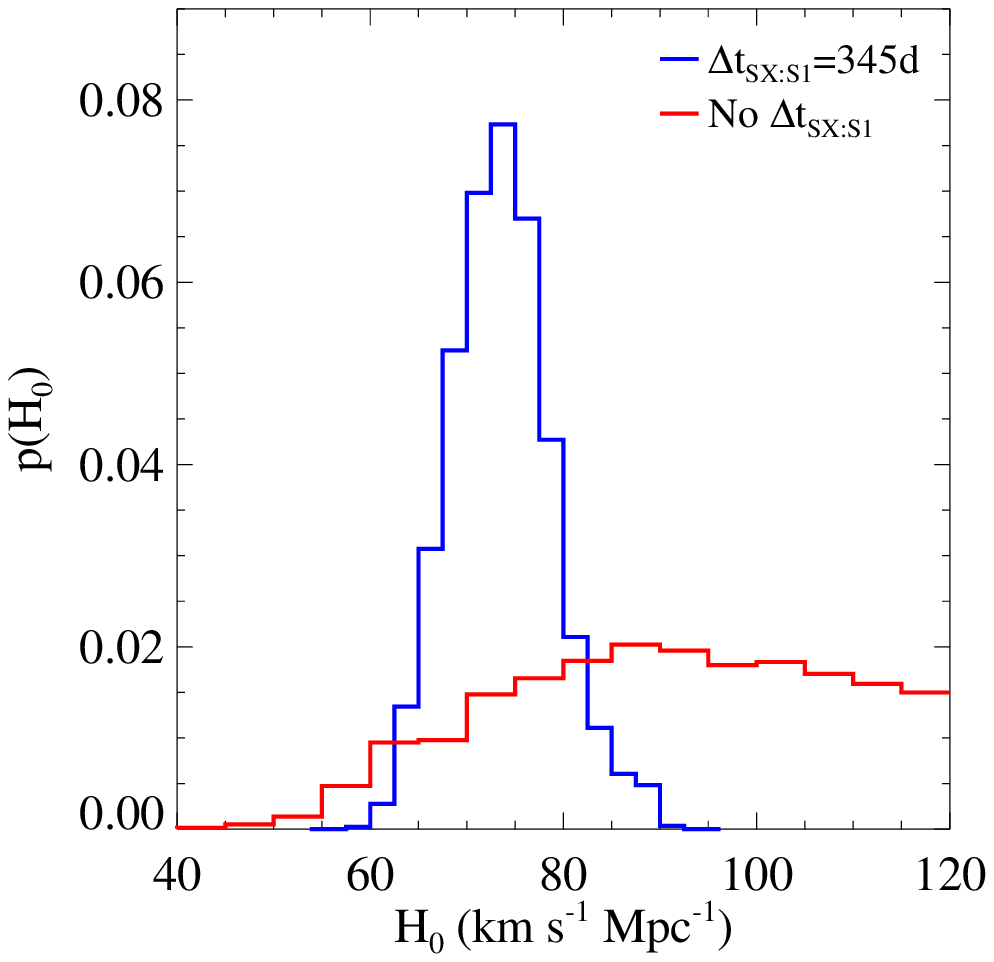}
\includegraphics[width=0.4\textwidth]{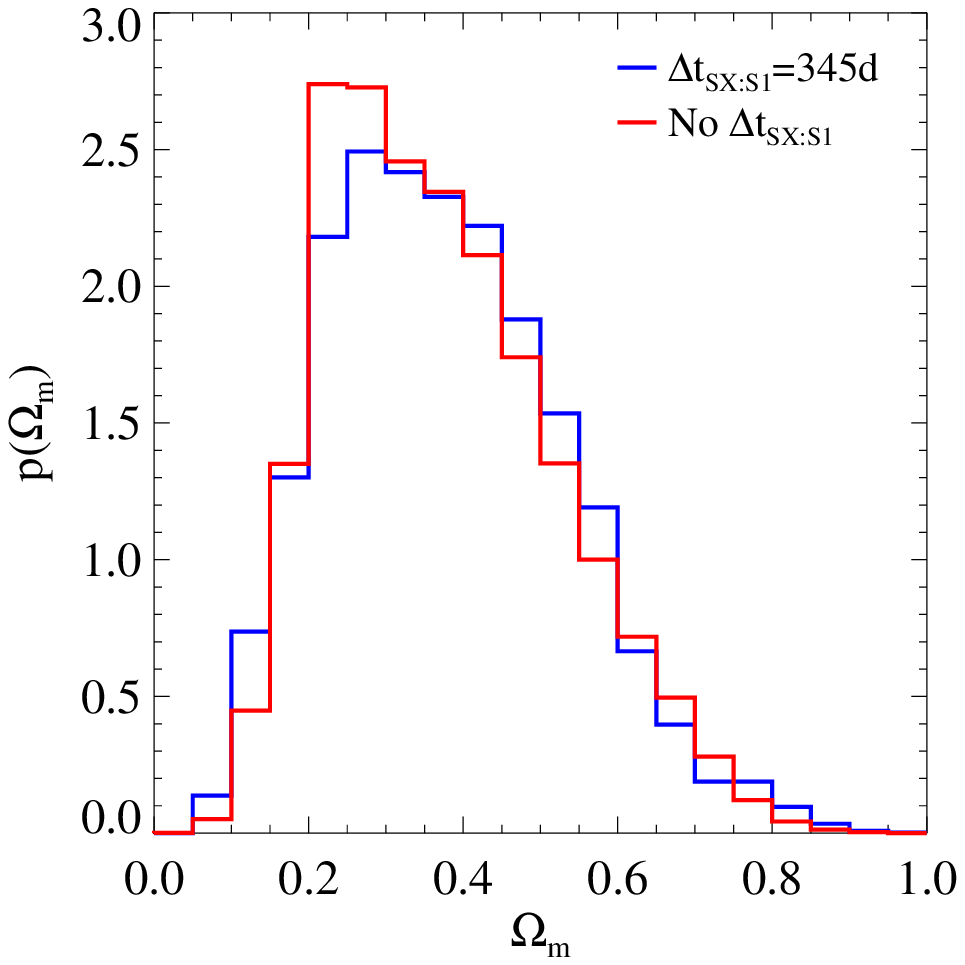}
\caption{Marginalized probability distribution functions of $H_{0}$ (on the left) and $\Omega_{\rm m}$ (on the right) for reference models with a time delay between SX and S1 of $345 \pm 10$ days (blue) and no time delay (red). Flat $\Lambda$CDM models ($\Omega_{\rm m}+\Omega_{\Lambda} = 1$) with uniform priors on the values of the cosmological parameters ($H_{0} \in [20,120]$ km~s$^{-1}$~Mpc$^{-1}$ and $\Omega_{\rm m} \in [0,1]$) are considered. Final MCMC chains have $2 \times 10^{6}$ samples for each model.}
\label{fig3}
\end{figure*}

In Figure \ref{fig3}, we illustrate explicitly the relevance of the longest measurable time-delay value between the multiple images of SN Refsdal, i.e. the SX-S1 time delay, to the cosmological inference within our reference model. We contrast the inferred results for $H_{0}$ and $\Omega_{\rm m}$ with a $345 \pm 10$ days time delay and without this time delay. We obtain that the probability distribution function of $H_{0}$ changes dramatically, from unimodal with a well-defined peak and small dispersion to almost flat over the largest values considered within the uniform prior. Remarkably, the probability distribution function of $\Omega_{\rm m}$ is instead essentially unaltered, with very similar median values and (1$\sigma$) statistical errors.

\begin{figure}
\centering
\includegraphics[width=0.49\textwidth]{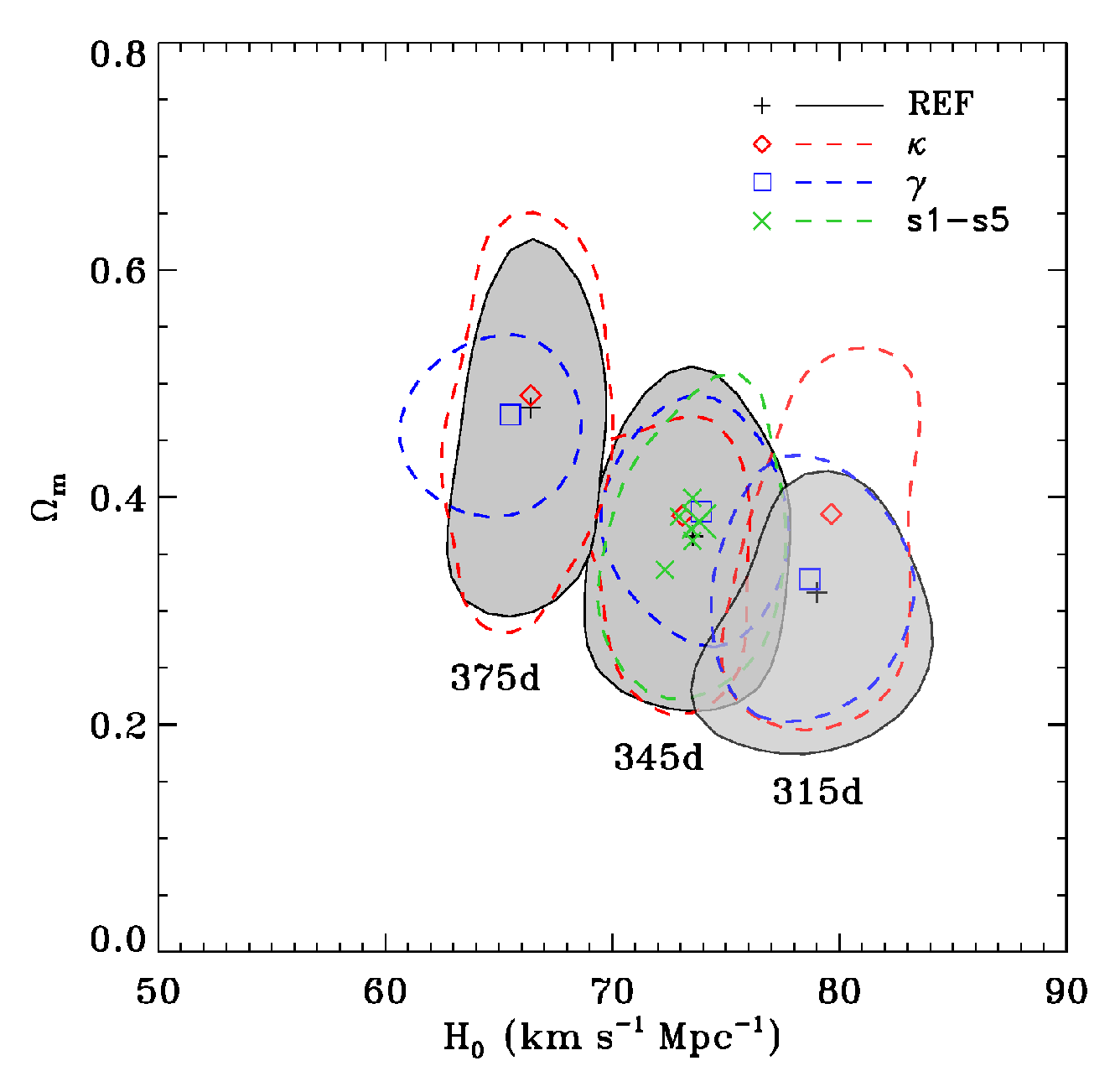}
\caption{Median values and contour levels, at 1$\sigma$ CL, of $H_{0}$ and $\Omega_{\rm m}$ for different lensing models. Time delays between SX and S1 of $315 \pm 10$, $345 \pm 10$, and $375 \pm 10$ days are adopted. Reference (REF/r) models are shown with black pluses and solid contours filled in gray. The addition of a constant sheet of mass ($\kappa$) at the cluster redshift (diamonds and dashed red contours), the use of a variable central density slope ($\gamma$) for the cluster main halo (squares and dashed blue contours), and $10\%$ scatter (s) on the values of $\vartheta_{\mathrm{E}}$ and $r_{\mathrm{t}}$ of the cluster members (crosses and dashed green averaged contours) are tested. Flat $\Lambda$CDM models ($\Omega_{\rm m}+\Omega_{\Lambda} = 1$) with uniform priors on the values of the cosmological parameters ($H_{0} \in [20,120]$ km~s$^{-1}$~Mpc$^{-1}$ and $\Omega_{\rm m} \in [0,1]$) are considered. Final MCMC chains have $2 \times 10^{6}$ samples for each model.}
\label{fig1}
\end{figure}

\begin{figure*}
\centering
\includegraphics[width=0.80\textwidth]{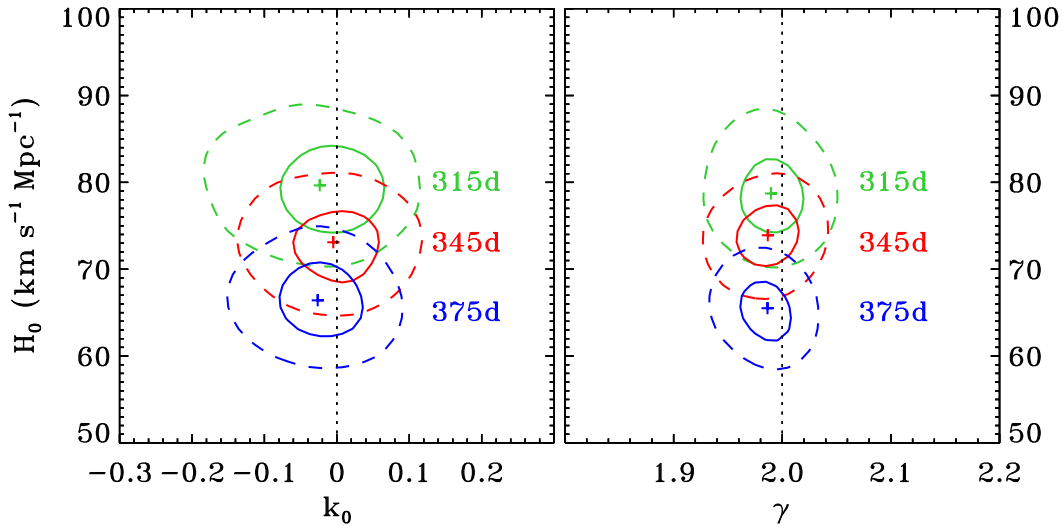}
\caption{Median values (pluses) and contour levels, at 1 (solid) and 2 (dashed) $\sigma$ CL, of $H_{0}$, k$_{0}$ (value of the convergence of a constant sheet of mass at the cluster redshift) and $\gamma$ (value of the slope of a power-law profile for the mass density of the cluster main halo) for different ($\kappa$ and $\gamma$) lensing models. Time delays between SX and S1 of $315 \pm 10$ (green), $345 \pm 10$ (red), and $375 \pm 10$ (blue) days are adopted. The fixed values of k$_{0}=0$ and $\gamma=2$ used in the reference (r) models are shown with vertical dotted lines. Flat $\Lambda$CDM models ($\Omega_{\rm m}+\Omega_{\Lambda} = 1$) with uniform priors on the values of the cosmological parameters ($H_{0} \in [20,120]$ km~s$^{-1}$~Mpc$^{-1}$ and $\Omega_{\rm m} \in [0,1]$) and on the values of k$_{0}$ ($\in [-0.2,0.2]$) and $\gamma$ ($\in [1.4,2.6]$) are considered. Final MCMC chains have $2 \times 10^{6}$ samples for each model.}
\label{fig1b}
\end{figure*}

\begin{figure*}
\centering
\includegraphics[height=0.98\textwidth,angle=270]{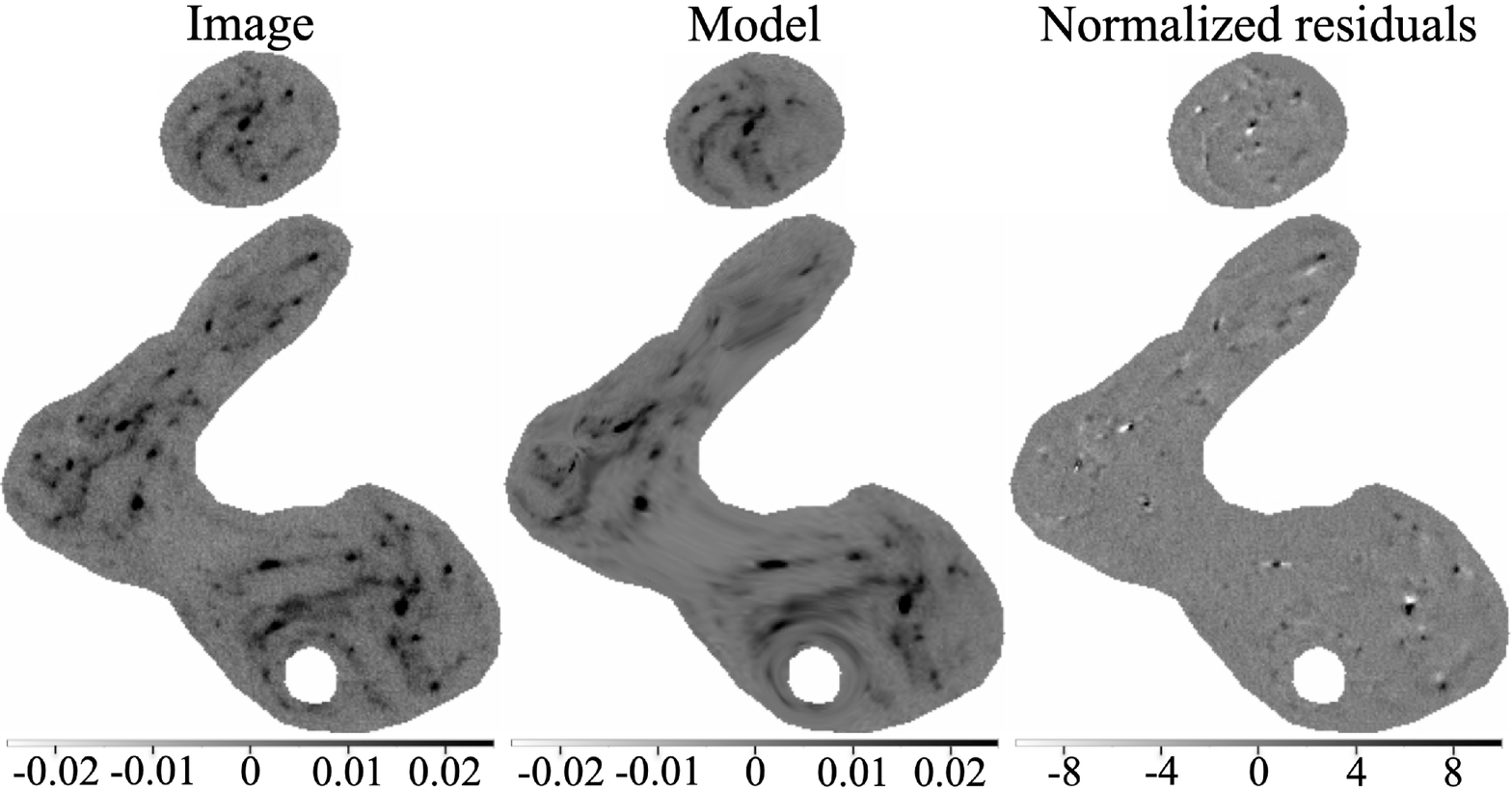}
\caption{Observed (on the left), model-predicted (in the middle) and normalized residual (on the right) images of the surface brightness distributions of the SN Refsdal host for the reference model with a time delay between SX and S1 of $345 \pm 10$ days. The original data (in units of counts per second) is the linear combination of the \HST\ F606W and F435W final images from the HFF project, optimized to suppress the flux contamination by the cluster member galaxies. The images have a pixel size of 0.06\arcsec. The source surface brightness is reconstructed on a $75 \times 75$ pixel grid.}
\label{fig5}
\end{figure*}

\begin{figure}
\centering
\includegraphics[width=0.49\textwidth]{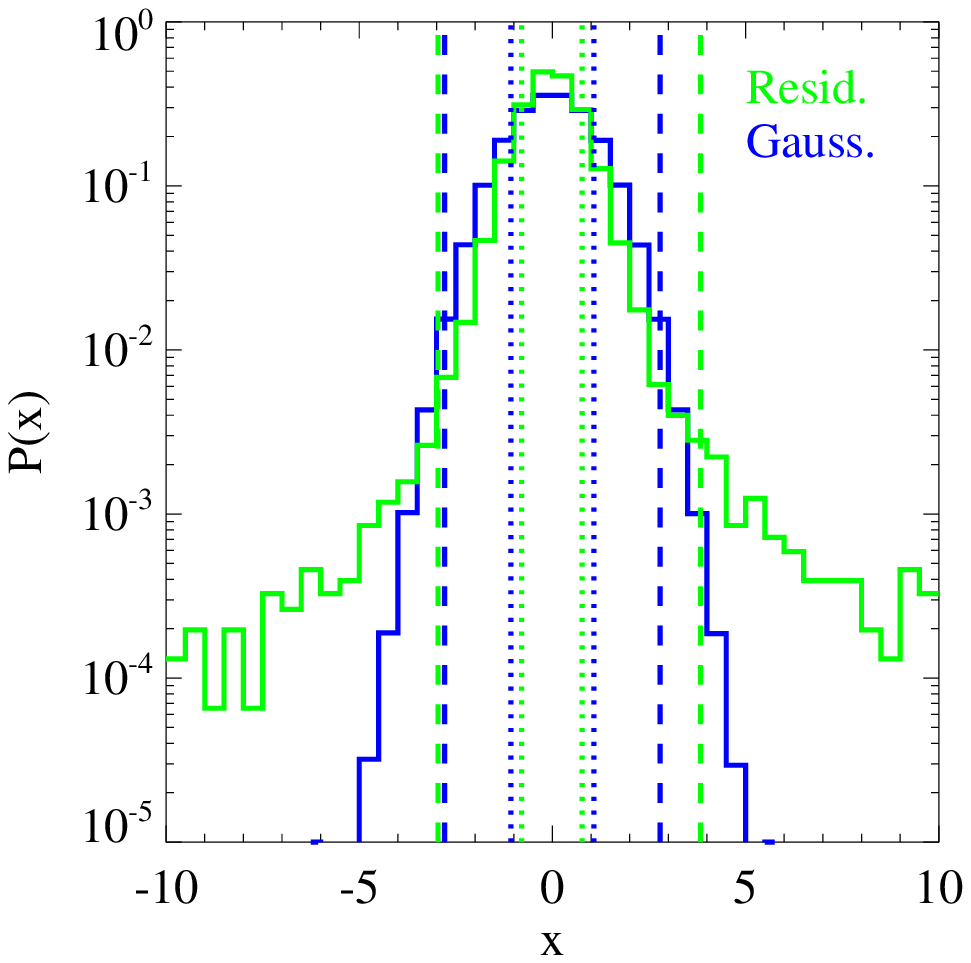}
\caption{Probability distribution functions of the normalized residuals (green) between the observed and model-predicted surface brightness distributions of the SN Refsdal host for the reference model with a time delay between SX and S1 of $345 \pm 10$ days (see the right panel in Figure \ref{fig5}) and of a normal distribution (blue) with standard deviation equal to that of the normalized residuals. Vertical dotted and dashed lines represent, respectively, the 68\% and 99\% probability intervals.}
\label{fig4}
\end{figure}

We summarize in Table \ref{stat} and in Figures \ref{fig1} and \ref{fig1b} the outcomes of our tests to estimate the systematic uncertainties on the values of $H_{0}$ and $\Omega_{\rm m}$, as detailed in Sect.~\ref{sec:met}. The most important result is that, for a fixed value of the SX-S1 time delay, the inferred values of the cosmological parameters are not significantly affected by the modeling details (considering their errors). For instance, when we use a time-delay value of $345 \pm 10$ days for ${\rm \Delta t_{\rm SX:S1}}$, the median values of $H_{0}$ and $\Omega_{\rm m}$ vary, respectively, between 72.9 and 73.9 km~s$^{-1}$~Mpc$^{-1}$ and 0.34 and 0.40 and the results of our reference model are not biased low or high. The combination of the eight final MCMC chains of the models r, $\kappa$, $\gamma$, and s1-5 with ${\rm \Delta t_{\rm SX:S1}} = 345 \pm 10$ days provides a global MCMC chain with $1.6 \times 10^{7}$ samples with median values and 1$\sigma$ uncertainties of 73.3 km s$^{-1}$ Mpc$^{-1}$ and 0.37 and 6\% and 40\% for $H_{0}$ and $\Omega_{\rm m}$, respectively. By comparing these results with those published in G18, we can conclude that the error budget on the estimate of $H_{0}$ is strikingly dominated by the statistical uncertainties and the known systematics play only a secondary role. We notice that the latter are slightly more relevant for the measurement of $\Omega_{\rm m}$.

We remark that very similar general conclusions hold when the other two values, i.e. 315 and 375 $\pm 10$ days, of the SX-S1 time delay are used. As discussed above, the inferred median values of $H_{0}$ decrease on average with increasing values of ${\rm \Delta t_{\rm SX:S1}}$. The results of these tests might also suggest that higher values of $\Omega_{\rm m}$ are preferred by higher values of ${\rm \Delta t_{\rm SX:S1}}$. Moreover, the 1$\sigma$ errors of $H_{0}$ and $\Omega_{\rm m}$ decrease mildly with increasing values of the SX-S1 time delay. This reflects the decreasing relative errors of the assumed constant 10 day uncertainty on longer time delays.

The juxtaposition of the observed and model-predicted surface brightness distributions of the SN Refsdal host shown in Figure \ref{fig5}, similarly to the comparison shown in Figure 7 by G16, strongly supports the validity of our results. Over the selected regions, composed of more than $3 \times 10^{4}$ \HST\ pixels, the mean and standard deviation values of the normalized residuals (i.e., the difference between the observed and model-predicted surface brightness values divided by the observational uncertainties) are, respectively, equal to $-0.002$ and $1.08$. This means that the multiple images of the SN Refsdal host, with their many star-forming regions or knots, are very well reconstructed, both in their positions and fluxes. In Figure \ref{fig4}, we show the probability distribution of our normalized residuals and that of a normal distribution with the same standard deviation. We notice that the former has an almost perfectly symmetric shape and we measure that 78.5\% (96.5\%) of the total number of pixels have normalized residuals smaller than 1 (2), to be compared to the 68.3\% (95.5\%) probability of a standard normal distribution. The fraction of pixels with normalized residuals larger than 5 is 0.5\% ($6 \times 10^{-5}$\% for a standard normal distribution). We have checked that the vast majority of these pixels are located around the SN host galaxy's peaks of emission, where the finite resolution of the grid on the source plane, remaining small offsets between the observed and model-predicted image positions, and the so-far neglected contribution of the point spread function make accurate image reconstructions more difficult to achieve. Considering the relatively simple subtraction of the cluster member flux (that can still be improved), the residuals at the positions of the S1-S4 and SX multiple images of SN Refsdal are remarkably small.

\begin{figure*}
\centering
\includegraphics[width=0.80\textwidth]{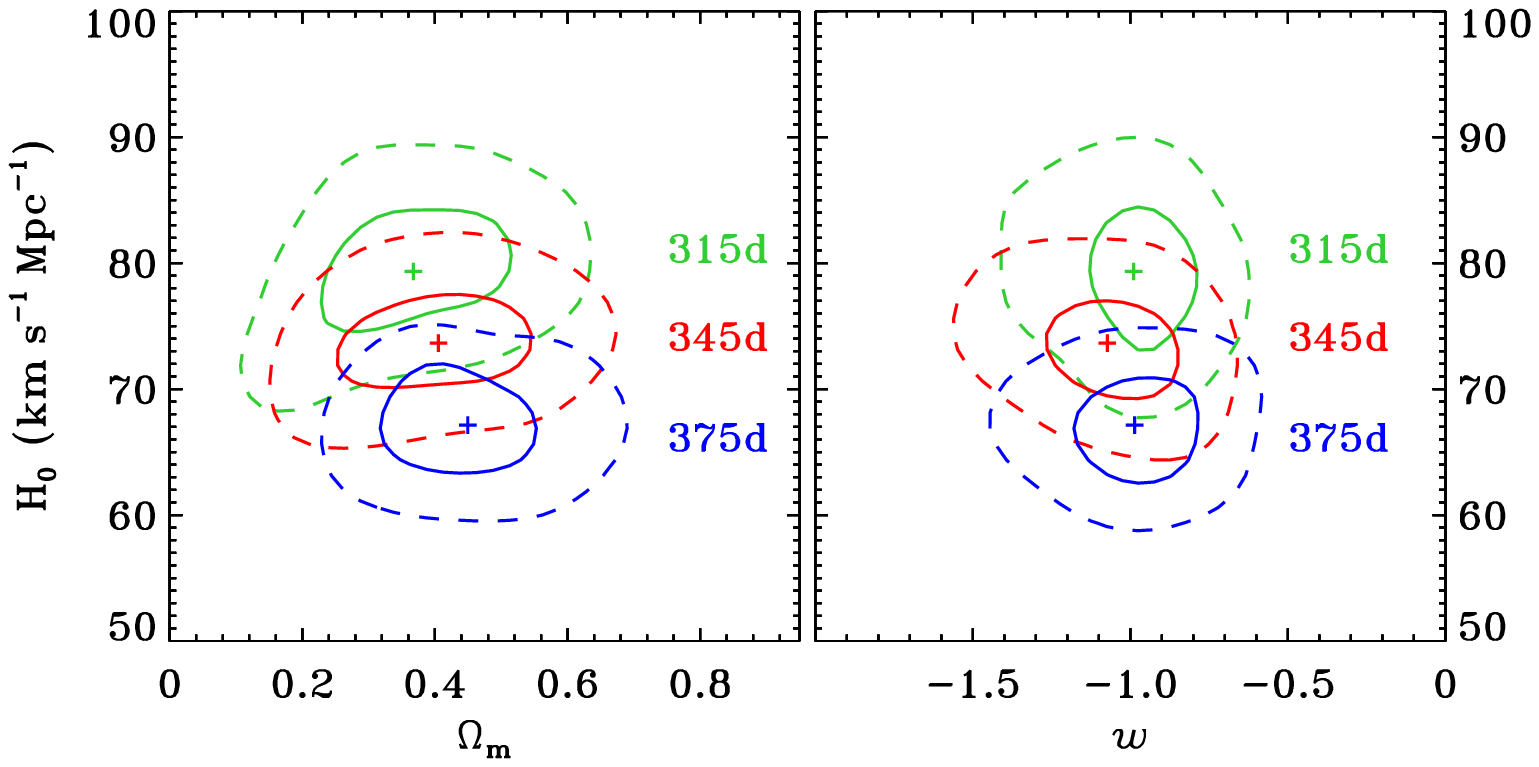}
\caption{Median values (pluses) and contour levels, at 1 (solid) and 2 (dashed) $\sigma$ CL, of $H_{0}$, $\Omega_{\rm m}$ and $w$ for the reference (r) models. Time delays between SX and S1 of $315 \pm 10$ (green), $345 \pm 10$ (red), and $375 \pm 10$ (blue) days are adopted. Flat $w$CDM models ($\Omega_{\rm m}+\Omega_{\Lambda} = 1$) with uniform priors on the values of the cosmological parameters ($H_{0} \in [20,120]$ km~s$^{-1}$~Mpc$^{-1}$, $\Omega_{\rm m} \in [0,1]$ and $w \in [-2,0]$) are considered. Final MCMC chains have more than $1.2 \times 10^{6}$ samples for each model.}
\label{fig6}
\end{figure*}
 
Finally, in Figure \ref{fig6} we show the inference on the values of the cosmological parameters $H_{0}$, $\Omega_{\rm m}$ and $w$. If we consider our reference model (with ${\rm \Delta t_{\rm SX:S1}} = 345 \pm 10$ days) and vary at the same time all the strong lensing model parameters, we obtain the following noteworthy 1$\sigma$ CL constraints: $73.6^{+4.6}_{-4.1}$ km s$^{-1}$ Mpc$^{-1}$ for $H_{0}$, $0.41^{+0.15}_{-0.15}$ for $\Omega_{\rm m}$ and $-1.07^{+0.22}_{-0.27}$ for $w$. The achieved precision on the
values of $\Omega_{\rm m}$ and $w$ is linked to the combined information coming from the time delays and the positions of the multiple images of sources at different redshifts.

\section{Discussion}
\label{sec:disc}

From the results presented in the previous section, it is clear that the most accurate and precise measurements of the values of the cosmological parameters from MACS 1149 and SN Refsdal will require an additional run of the strong lensing models, that include the final measurement of the value and uncertainty of the SX-S1 time delay. Although some scaling relations between the value of ${\rm \Delta t_{\rm SX:S1}}$ and those of the cosmological parameters exist (as highlighted in G18 and above), the best option to infer the exact values of $H_{0}$ and $\Omega_{\rm m}$ and their errors, that aim to be competitive with the most recent ones from other cosmological probes, is definitely not through a simple interpolation of the results already obtained with the current plausible but approximate time delay. We have also demonstrated that a precise measurement of the value of $H_{0}$ is not achievable without some information on ${\rm \Delta t_{\rm SX:S1}}$.

By comparing different results of the strong lens time delay method, with SN Refsdal in MACS 1149 (as shown here) and with lensed quasars in the galaxy-scale systems of the H0LiCOW program (\citealt{suy17}), we can conclude that (1) the relative error on the inferred value of $H_{0}$ from a single (galaxy or cluster) strong lensing system is similar (mean value of 6.4\% in Figure 2 of \citealt{won19}), (2) in a single lens cluster, we have the additional interesting possibility of estimating the value of $\Omega_{\rm m}$ (and $w$), thanks to the observations of several multiple-image families with spectroscopically confirmed redshifts and to the measurement of the time-delay values between the multiple images of time-varying sources, and (3) the observed positions of many spectroscopic multiple images (some of which are crucial to constraining the lens tangential and radial critical curves at different redshifts) provide precise calibrations of all the different mass components (i.e., extended dark-matter halos, cluster members, and possibly hot gas) included in the modeling of a galaxy cluster and, thus, also a good approximation of the effect of the ``environment'' in the regions adjacent to where the time delays are measured. This is supported by the almost negligible impact of the inclusion in the ($\kappa$) models of a constant sheet of mass at the lens redshift (Figure \ref{fig1}). In particular, as shown in Figure \ref{fig1b}, the inferred value of k$_{0}$ is approximately 0 and not correlated with that of $H_{0}$, whereas this is not the case when only multiple images at a single redshift (i.e., that of SN Refsdal) are used (see Appendix \ref{sec:appA}).

Finally, we notice that relatively large residuals have already been reported in close proximity to the most luminous pixels of multiply imaged QSOs with measured time delays in the image reconstruction of the lensed quasar and host galaxy surface brightness distributions (e.g., \citealt{won17}).
As in the case of SN Refsdal host, this can be ascribed mainly to the finite grid resolution on the source plane. We suggest that a comparison between the observed and model-predicted surface brightness distributions of SN Refsdal host, as that shown in Figures \ref{fig5} and \ref{fig4}, should be used to test the goodness of any strong lensing model which provides estimates of the values of cosmological parameters from the time delays of the SN Refsdal multiple images. More in general, we believe that the extended surface brightness modeling of multiply imaged QSO and SN hosts, although computationally expensive, will help to make progress in time-delay cosmography with lens galaxy clusters, as already demonstrated with lens galaxies.

\section{Conclusions}
\label{sec:conc}

We have shown that accurate and precise estimates of the value of the Hubble constant can be obtained in lens galaxy clusters, when a large set of spectroscopically confirmed multiple images of sources at different redshifts and the time delays of the multiple images of a time-varying source are measured. In flat $\Lambda$CDM models, we have demonstrated that a full and blind strong lensing analysis of the HFF galaxy cluster MACS J1149.5$+$2223, with an error of 10 days (i.e., $\sim$3\%) on the longest measurable time delay between the multiple images of SN Refsdal provides approximately 6\% and 40\% total uncertainties for $H_{0}$ and $\Omega_{\rm m}$, respectively.
A range of values (315-375 days) for this time delay has been explored, since its accurate measurement is still not known at the time of analysis and writing. We have tested several possible sources of systematic uncertainties and confirmed that their contribution to the estimated values of the cosmological parameters is significantly less important than that of the statistical uncertainties. We have verified that the effect of the ``mass-sheet degeneracy'' is considerably reduced, thanks to the presence of tens of multiply imaged sources at different distances behind the lens. We have proved that the reconstructed extended surface brightness distribution of the SN host validates the reliability of our models. When the values of $H_{0}$ and $\Omega_{\rm m}$ are measured, we remark the importance of considering the full covariance between all the parameters defining the cosmological model and the total mass distribution of the cluster lens. We conclude that time-delay cluster lensing will turn into a new valuable cosmological tool, once more high-quality data of cluster strong lenses become available from the next deep and wide surveys.




\smallskip

\acknowledgments
C.G. and P.R. acknowledge support through grant MIUR2017 WSCC32 and C.G. also through grant no.~10123 of the VILLUM FONDEN Young Investigator Programme.
S.H.S.~thanks the Max Planck Society for support through the Max Planck Research Group.
A.M. acknowledges funding from the INAF PRIN-SKA 2017 program 1.05.01.88.04.
This work is based in large part on data collected in service mode at ESO VLT, through the Director's Discretionary Time Programme 294.A-5032.
The CLASH Multi-Cycle Treasury Program is based on observations made with the NASA/ESA {\it Hubble Space Telescope}. The Space Telescope Science Institute is operated by the Association of Universities for Research in Astronomy, Inc., under NASA contract NAS 5-26555. ACS was developed under NASA Contract NAS 5-32864. 

\appendix

\section{The mass-sheet degeneracy}
\label{sec:appA}

\begin{figure}
\centering
\includegraphics[width=0.49\textwidth]{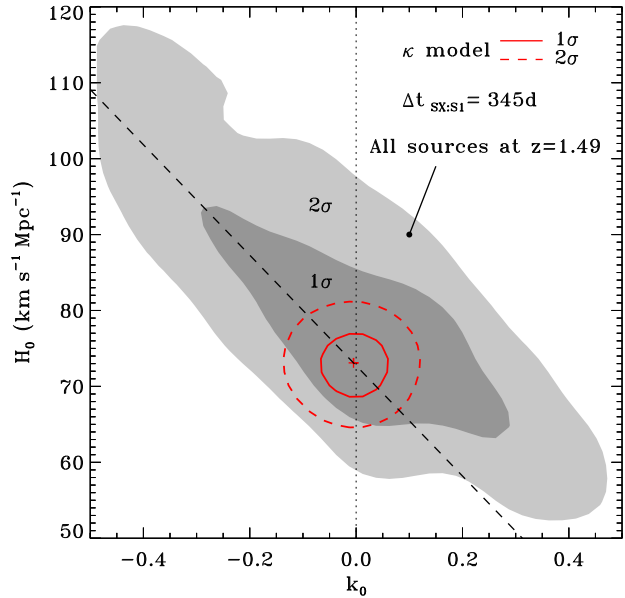}
\caption{Median values (plus) and contour levels, at 1 and 2 $\sigma$ CL, of $H_{0}$ and k$_{0}$ for the $\kappa$ lensing models with all (89) multiple images (red) and with only those (63) belonging to SN Refsdal and its host (gray). A time delay between SX and S1 of $345 \pm 10$ days is adopted. The fixed value of k$_{0}=0$ used in the reference (r) models is shown with the vertical dotted line. The dashed line illustrates the theoretical effect of the so-called ``mass-sheet degeneracy'' (\citealt{sch13}). Flat $\Lambda$CDM models ($\Omega_{\rm m}+\Omega_{\Lambda} = 1$) with uniform priors on the values of the cosmological parameters ($H_{0} \in [20,120]$ km~s$^{-1}$~Mpc$^{-1}$ and $\Omega_{\rm m} \in [0,1]$) and on the value of k$_{0}$ ($\in [-0.2,0.2]$ or $[-0.5,0.5]$) are considered. Final MCMC chains have $2 \times 10^{6}$ samples for each model.}
\label{fig7}
\end{figure}

We compare here the results of two ($\kappa$) models with a constant sheet of mass at the cluster redshift and with ${\rm \Delta t_{\rm SX:S1}} = 345 \pm 10$ days, where we consider either the entire sample of 89 multiple images from 28 sources at different redshifts (as presented and discussed above) or only the 63 multiple images from SN Refsdal and its host, all at $z = 1.489$. In Figure \ref{fig7}, we show the inferred values of $H_{0}$ and k$_{0}$. We find that the use of many multiple images belonging to sources located at different distances reduces significantly the effect of the so-called ``mass-sheet degeneracy'' (\citealt{fal85}; \citealt{sch19}). At the 1$\sigma$ CL, the credible interval of k$_{0}$ varies from [$-$0.30,0.24]
 (extending over a large fraction of the adopted uniform
 prior, [−0.5,0.5], chosen here to illustrate better the parameter degeneracy), when sources at a single redshift are included in the model, to [$-$0.08,0.06], when all sources at different redshifts are considered. This translates into an approximately $9\%$ difference in the median value of $H_{0}$ (and of $\Omega_{\rm m}$) and in a remarkable reduction by a factor of more than 3, from $\sim$21\% to $\sim$6\%, for its uncertainty (from $\sim$63\% to $\sim$40\% for the uncertainty on $\Omega_{\rm m}$).

\end{document}